\documentclass[12pt,preprint]{aastex}

\slugcomment{\today}

\def\ea{{\it et al.\,}}
\def\be{\begin{equation}}
\def\ee{\end{equation}}

\def\bs{\bigskip}

\def\np{\newpage}

\begin{document}
\title{Redshift Dependence of the CMB Temperature from S-Z Measurements}
\bs
\bs

\author{G. Luzzi\altaffilmark{1}, M. Shimon\altaffilmark{2}, L. Lamagna\altaffilmark{1}, Y. Rephaeli\altaffilmark{2,3}, M. De Petris\altaffilmark{1},
A. Conte\altaffilmark{1}, S. De Gregori\altaffilmark{1}, E.S. Battistelli\altaffilmark{1}}
\altaffiltext{1}{Department of Physics, University ``La
Sapienza'', P.le A. Moro 2, 00185, Rome,
Italy}
\altaffiltext{2}{Center for Astrophysics and Space Sciences,
University of California, San Diego,9500 Gilman Drive, La Jolla,
CA, 92093-0424}
\altaffiltext{3}{School of Physics and Astronomy,
Tel Aviv University, Tel Aviv, 69978, Israel}

\begin{abstract}
We have determined the CMB temperature, $T(z)$, at redshifts in
the range 0.023-0.546, from multi-frequency measurements of the
S-Z effect towards 13 clusters. We extract the parameter $\alpha$
in the redshift scaling $T(z)=T_{0}(1+z)^{1-\alpha}$, which
contrasts the prediction of the standard model ($\alpha=0$) with
that in non-adiabatic evolution conjectured in some alternative
cosmological models. The statistical analysis is based on two main
approaches: using ratios of the S-Z intensity change, $\Delta I$,
thus taking advantage of the weak dependence of the ratios on IC
gas properties, and using directly the $\Delta I$ measurements. In
the former method dependence on the Thomson optical depth and gas
temperature is only second order in these quantities. In the
second method we marginalize over these quantities which appear to
first order in the intensity change. The marginalization itself is
done in two ways - by direct integrations, and by a Monte Carlo
Markov Chain approach. Employing these different methods we obtain
two sets of results that are consistent with $\alpha=0$, in
agreement with the prediction of the standard model.

\end{abstract}


\section{Introduction}

The variation of the Cosmic Microwave Background (CMB) temperature
with redshift is a basic relation which in adiabatically evolving
cosmological models is $T(z) = T_{0}(1+z)$, normalized to the
COBE/FIRAS value at the present epoch, $T_{0}=2.725\pm 0.002$K
\citet{mather1999}. In light of the fundamental role the CMB plays
in cosmology, and given our detailed knowledge of its spatial
structure, the lack of precise observational confirmation of this
important relation is rather surprising. Further motivation to
measure $T(z)$ is provided by the prediction of different
non-linear redshift scaling law in alternative cosmological models
(e.g. \citealt{overduin1998, matyjasek1995, puy2004}). In
particular, we consider here the scaling law proposed by
\cite{lima2000}, $T(z)=T_{0}(1+z)^{1-\alpha}$, with $\alpha$ a
free (constant) parameter; this scaling is presumed to be a
consequence of photon number and radiation entropy
non-conservation.

The $T(z)$ dependence has traditionally been tested by
measurements of quasar absorption spectra which include
fine-structure lines from interstellar atoms and ions (e.g.,
\cite{losecco2001}). Using this method the CMB temperature was
measured to a maximum redshift of $3.025$, at which $T\simeq
12.6^{+1.7}_{-3.2}$ K was determined from an analysis of the CII
fine-structure lines in the damped $Ly_{\alpha}$ system towards
the quasar Q0347--3819 \cite{molaro2002}. The fact that the CMB is
not the only radiation field populating the energy levels (from
which the transitions occur), and lack of detailed knowledge of
the physical conditions in the absorbing clouds - is a major
source of systematic uncertainty (e.g., \citealt{combes1999}). A
strong limit on $\alpha$ was deduced \cite{opher2005} using CMB
and galaxy distribution data. However, this method cannot be used
to determine the $T(z)$ scaling.

The possibility of determining $T(z)$ from measurements of the
Sunyaev-Zeldovich (S-Z) effect had been suggested long ago
\cite{fabbri1978, reph1980}. The effect - Compton scattering of
the CMB by hot intracluster (IC) gas - is a small change of the
CMB spectral intensity, $\Delta I$, which depends on the
integrated IC gas pressure along the line of sight (los) to the
cluster. The steep frequency dependence of the change in the CMB
spectral intensity, $\Delta I$, due to the S-Z effect allows the
CMB temperature to be estimated at the redshift of the cluster.
Since the ratio of the values of $\Delta I$ measured at two
frequencies is essentially independent of the cluster properties,
the value of the temperature at the cluster redshift can be
deduced directly from this ratio \cite{reph1980}. The much
improved capability of precise multi-frequency measurements of the
effect enhances interest in this method to measure $T(z)$ in
nearby and moderately distant clusters.

The first attempt to determine $T(z)$ from an analysis of
multifrequency S-Z measurements in the Coma and A2163 clusters of
galaxies was reported in \cite{battiste2002}. Preliminary results
from an analysis of a larger sample of clusters, with
$z=0.023-0.546$, was presented in \cite{degregori2008}. In this
paper we report the final results from the current cluster sample
carrying out a significantly expanded statistical data analysis.
The statistical analysis is based on both ratios of $\Delta I$,
which are weakly dependent on the cluster properties, as well as
the values of the individual $\Delta I$ whose dependence on
cluster properties is marginalized over (also) the Thomson optical
depth. We discuss the consistency between these two analysis
methods.

In Section \ref{sec:method} we outline the basic methods used to
determine $T(z)$. Data analysis and the results for $\alpha$ are
presented in Section \ref{sec:dataanalysis}. We end with a brief
summary in Section \ref{sec:summary}.

\section{Method}\label{sec:method}

Compton scattering of the CMB in a cluster causes a change of
intensity that can be written as \be\label{eq:deltaISZ} \Delta I={
2 k^{3} T^{3} \over h^{2}c^{2} } {x^4e^x \over (e^x -1)^2} \int
d\tau \biggr[\theta f_{1}(x) - \beta + R(x, \theta, \beta) \biggl
] \, , \ee where $x\equiv h\nu /kT$ is the dimensionless
frequency, $\theta= kT_e /mc^2$ is the electron temperature in
units of the electron rest energy, and $\beta$ is the line of
sight (los) component of the cluster velocity (divided by $c$) in
the CMB frame. The integral is over the Thomson optical depth,
$\tau$. Both the thermal \cite{SunZel72} and kinematic
\cite{SunZel80} components of the effect are included in Equation
(1) in the first two terms, and in the function
$R(x,\theta,\beta)$. This latter term is the relativistic
correction \cite{reph95} to the non-relativistic expressions for
the thermal and kinematic components. An analytic approximation
(which is sufficiently exact even close to the crossover
frequency) can be written in the form \be\label{eq:relcorr}
R(x,\theta, \beta) \simeq \theta^{2} \biggr[f_{2}(x)+\theta
f_{3}(x) + \theta^{2}f_{4}(x) + \theta^{3}f_{5}(x) \biggl] - \beta
\theta \biggr[ g_{1}(x) + \theta g_{2}(x) \biggl] +
\beta^{2}\biggr[1 + \theta g_{3}(x) \biggl] \,. \ee The spectral
functions $f_{i}$ and $g_{i}$ are specified in \cite{shimon2004};
see also \cite{itoh2002}.

With $T(z) = T(0)(1+z)^{(1-\alpha)}$, and $\nu = \nu_{0}(1+z)$,
the non-dimensional frequency depends on redshift if $\alpha\neq
0$, $x = x_{0}(1+z)^{\alpha}$, with $x_{0}= h\nu_{0}/kT(0)$. The
steep frequency dependence of the change in the CMB spectral
intensity allows the CMB temperature to be estimated at the
redshift of the cluster. In the non-relativistic limit $\Delta I$
depends linearly on the Comptonization parameter, $y = \int \theta
d\tau$, which includes all dependence on IC gas properties. A
ratio of values of $\Delta I$ at two frequencies is then
essentially independent of the cluster properties. In the more
general case, the first term in the square parentheses in Eq.
(\ref{eq:deltaISZ}) still dominates over the other two, except near
the crossover frequency, where the sum of the temperature
dependent terms vanishes. For values of $x$ outside some range
(roughly, $3.5 < x < 4.5$), the dependence of $\Delta I$ on
$\beta$ is very weak since the observed temperature range in
clusters corresponds to $0.006 < \theta < 0.03$, whereas typically
$\beta < 0.002$. Therefore, data points at or near the crossover
frequency are noise-dominated, and thus carry relatively low
statistical weight in the likelihood function we construct for
$\alpha$ or $T(z)$. Nevertheless, we include these points in our
analysis.

\section{Data Analysis}\label{sec:dataanalysis}

We analyzed results of multi-frequency S-Z measurements toward 13
clusters spanning the redshift range $0.023-0.550$. The dataset
includes measurements with the BIMA, OVRO, SUZIE-II, SCUBA, and
MITO telescopes, as well as gas temperatures determined from X-ray
measurements. We assumed a gaussian profile for the spectral bands
of each experiment. In order to evaluate systematics induced by
different spectral efficiencies of the bands, we repeated the
analysis with squared profiles. The error contribution is
negligible as compared with uncertainties in values of the central
frequency and bandwidth. The clusters, redshifts, and gas
temperatures are listed in Table \ref{tab:tableX}, and the S-Z
data are in Table \ref{tab:tableSZ}.
In the analysis we allowed for calibration uncertainty, considered
as an uncertain scale factor, which was modeled as a gaussian with
mean 1 and $10\%$ standard deviation
%\footnote{The calibration errors for the experiments considered
%are at a maximum level of $10\%$, based on observations of the
%brightness of planets. We have adopted a conservative approach in
%assuming maximum errors in order not to be biased by the different
%calibration procedures of the different experiments.}.
\footnote{The calibration errors for the experiments considered
are at a maximum level of $10\%$, based on observations of the
brightness of planets. We have adopted a conservative approach in
assuming maximum errors in order not to be biased by different
calibration procedures of different experiments.}.
This is an adequate approximation given the relatively narrow
widths of the respective gaussian distributions. For the radial
component of the cluster peculiar velocities we assume a universal
vanishing mean with a $1000$ km/s standard deviation.

The calculation begins with convolution of $\Delta I(\nu)$ with
the detector Gaussian spectral response function with width
$\sigma_{\nu_0}$, \be\label{eq:convolution} \Delta I(\nu_{0})
\equiv \frac{1}{\sqrt{2\pi}\sigma_{\nu_0}}
\int_{\nu=0}^{\infty}\Delta I(\nu)
e^{-\frac{(\nu-\nu_0)^{2}} {2\sigma_{\nu_0}^{2}}}d\nu \, , \ee
using the approximate analytic expression for (the
relativistically correct) $\Delta I(\nu)$ derived by
\cite{shimon2004}. Integration by parts then yielded an analytic
approximation to the frequency-convolved $\Delta I(\nu)$, accurate
to order $(\sigma_{\nu_0}/\nu)^{4}$. The sole purpose of deriving
the latter analytic (rather cumbersome) approximations (which are
not specified here) was to enable a faster convolution of the
theoretical expression for $\Delta I(\nu)$ with the detector
Gaussian spectral response.

The probability of measuring an intensity change $\Delta I_{obs}
(\nu_{j})$ at frequency $\nu_{j}$ towards a cluster at redshift
$z_{i}$, with velocity parameter $\beta_{i}$, gas parameters
$\tau_{i}$, $\theta_{i}$, and model parameter $\alpha$ is,
\begin{eqnarray}
P(\tau_{i},\theta_{i},\beta_{i};z_{i},\alpha)\propto\exp\{-\sum_{j}\frac{[\Delta
I(\tau_{i},\theta_{i},\beta_{i},\nu_{j};\alpha)-\Delta I_{{\rm
obs}}(\nu_{j})]^{2}} {2\sigma_{I}^{2}(\nu_{j})}\}
\end{eqnarray}
where $\Delta I$ is the result of convolution of the theoretically
predicted intensity change with the spectral response function
(calculated using Eq.\ref{eq:convolution}), and
$\sigma_{I}(\nu_{j})$ is measurement
error standard deviation.

The CMB temperature at the redshift of each cluster was first
extracted by performing statistical analysis on the ratios
$r_{ij}= \Delta T(\nu_i)/ \Delta T(\nu_j)$ ($\Delta T(\nu_i)$ is
the observed thermodynamic temperatures in the $i$-th photometric
band) marginalizing over $\theta$ and $\beta$ - the original
(hereafter) RI approach - and by using each of these separately
and marginalizing also over $\tau$, $\theta$ and $\beta$ - the DI
approach
\footnote{Note that for marginalizing over a range of gas
temperatures we have determined this range by using a beta profile
for the gas density. The modeling uncertainty due to the use of
this profile is negligible.}.
Assuming the S-Z data are gaussianly distributed allows
us to construct the relevant single-cluster and multiple-cluster
likelihood functions. We note that the high-dimensionality of the
marginalization, over the full parameter space, including also the
calibration uncertainty (see Section \ref{par:ratio}) can
potentially be a source for large numerical errors. We therefore
employed two separate pipelines in the DI analysis - a direct
%numerical integration and MCMC sampling.
numerical integration and Monte Carlo Markov Chain (MCMC) sampling.
After proper treatment of
the systematics the different methods give consistent results.

\subsection{Likelihood of Intensity Ratios}\label{par:ratio}

The RI approach - first employed in our previous work
\cite{battiste2002} - is based on the use of the likelihood
function of intensity ratios, thereby removing (to first order)
dependence on the Comptonization parameter \cite{reph1980}. By
doing so we largely avoid the need to account for modeling
uncertainties in the determination of the gas density and
temperature profiles (from X-ray surface brightness and spectral
measurements). Dependence on the measured gas temperatures, $T_e$,
is also largely removed in the leading term, although $T_e$ is
still needed to calculate the relativistic correction which
includes second order terms in $T_e$. Measurements at the
crossover frequency were not included in the analysis due to the
inherent problematics associated with inclusion of very weak
signals in the likelihood function (e.g., `Cauchy tail',
bimodality, etc), which are expected to be dominated by
instrumental noise and systematic uncertainties, so their relative
weight in the likelihood function would be negligible.

The distribution of ratios may have complicated features, but if
carefully applied can be a very powerful tool due to the reduced
number of essential parameters in the problem. In our case the
distribution of ratios of temperature changes is manifestly
non-gaussian, but we are interested in constraining the CMB
temperature (or the parameter $\alpha$) rather than $\Delta T$
itself. The distribution is expected to be normal when the cluster
sample is large. As shown in Appendix A, this approach has the
unwanted feature that the expectation value of the ratio is biased
to a degree which is of $O\left((\Delta T/T)^{2}\right)$, but in
the limit of very precise measurements the fractional error
decreases as does the degree of bias. This follows from the fact
that a uni-modal distribution function behaves as a gaussian
sufficiently close to its peak, and if measurement errors are
sufficiently small the function is sharply peaked. In other words,
when the function is highly concentrated near the peak, any value
of the parameter (for which the distribution function does not
vanish) is in the `gaussian-regime'. This guarantees that the
ratio-distribution is an unbiased gaussian. However, for the small
number of clusters in our present sample, and the low quality of
some of the measurements, our results are still affected by this
bias, but at a level smaller than the statistical uncertainty. The
situation will improve substantially in the near future, when
planned surveys with PLANCK and other S-Z projects will provide a
very large sample of clusters. A more complete discussion of the
ratio approach is given in Appendix A.

The joint distribution function is derived from a set of $n$ S-Z
measurements (at $n$ different frequency bands, excluding the
crossover frequency) of a given cluster. In doing so we choose the
S-Z data point with the smallest fractional error to be in the
denominator of the $n-1$ ratios so as to minimize the bias
(Appendix A). We construct the likelihood function of the
simultaneously calculated $n-1$ ratios for a given frequency (in
the denominator) obtaining a single likelihood function per
cluster.

The likelihood function that encapsulates the multi-cluster information
is defined as
%\begin{eqnarray}\label{eq:like-ratio}
%L(\alpha)\equiv\Pi_{i} L(C,\theta_{i}, \beta_{i};\alpha)
%\end{eqnarray}
\begin{eqnarray}\label{eq:like-ratio}
L(\alpha)\equiv\Pi_{i} L(\theta_{i}, \beta_{i}, C;\alpha)
\end{eqnarray}
where the index $i$ runs over all clusters in the sample. In the
following we marginalize over $\theta_i$, $\beta_i$, and
calibration uncertainty, denoted here by $C$. In the limit of
sufficiently large dataset, the joint likelihood function is
gaussian in $\alpha$ by virtue of the central limit theorem and in
accord with the reasoning we specified above. The average gas
temperatures and their errors, as obtained from X-ray
observations, are summarized in Table \ref{tab:tableX}. The
smaller (than thermal) kinematic component vanishes to first order
(due to motion either toward or away from the observer).

\subsection{The DI Approach}\label{sec:DIapproach}

As noted, we carried out another independent analysis based on
using the measured values of $\Delta I$ themselves, rather than
their ratios. This second approach was implemented in two separate
and very different treatments. The first consisted of the
construction of a likelihood function that incorporates the
multi-cluster information, with direct numerical 2D integrations
involved in the marginalization
%over $C$, $\tau$, $\theta$, and $\beta$
over $\tau$, $\theta$, $\beta$ and $C$
(the integration over $\beta$ and $C$ is carried out
analytically after Taylor expanding the integrand in powers of
$\beta$ and $C$)
%\begin{eqnarray}
%L(\alpha)\equiv\Pi_{i}L(C,\tau_{i},\theta_{i},\beta_{i};\alpha).
%\end{eqnarray}
\begin{eqnarray}
L(\alpha)\equiv\Pi_{i}L(\tau_{i},\theta_{i},\beta_{i},C;\alpha).
\end{eqnarray}
For $\tau$ we use a flat (positive definite) prior; for the other
parameters priors are chosen as in the RI approach (see Section
\ref{sec:dataanalysis}).

The second independent treatment is based on
%Monte Carlo Markov Chain (MCMC)
MCMC
sampling, which is very commonly used in a wide class
of problems, where complex dependencies between the parameters
greatly impact the likelihood function, so that direct
integrations could be quite tedious and prone to numerical
instabilities (e.g., \citealt{lewis2002, verde2007}). In this
treatment we construct a single likelihood for each cluster, in
order to get directly $T(z)$ at the cluster redshift, and also to
check parameter degeneracies.

The MCMC method constitutes a random sequence of realizations of
the fit parameters which - when properly sampled in parameter
space - tends to reproduce the posterior distribution for all the
parameters. The technique is easily implemented into an iterative
code that draws candidate sets of fitting parameters of chain
elements from a given proposal distribution, and inserts the candidates
into the chain if some proper statistical criterion is met. The sampling
approach we used is the one proposed by Metropolis and Hastings
\cite{metropolis1953, hastings1970}. According to this scheme,
each candidate set of parameters $\vartheta_i$ is drawn from a
gaussian centered over the previously accepted chain element
$\vartheta_{i-1}$. The width of the proposal distribution is a
critical parameter of the code which determines
the efficiency of the sampling algorithm (i.e. the capability to
sample from the multi-parameter posterior distribution in a given
number of realizations, or acceptance rate) and its ability to
explore the full domain of the fitting parameters. Different test
runs of the MCMC code were needed to find the best tradeoff
between acceptance rate and accuracy of the fit for given chain
lengths, by adjusting the width of the proposal distributions. In
the final configuration, the code was able to reconstruct the
posteriors with $5\%\div10\%$ acceptance rate and
sufficient samplings of the parameter space were achieved in less
than $10^6$ samplings. Convergence of the MCMC runs was also
tested through the Gelman-Rubin test \cite{gelman1992} which not
only tests convergence but can also diagnose poor mixing. We have
verified that the results are independent of the starting point in
parameter space by performing multiple shorter runs of the code
and confirming the statistical consistency of the results of the
single runs. Undersampling of the final chains was performed to
reduce the self-correlation which is naturally induced in a
sampled Markov Chain, and may therefore affect the estimate of
sample variances for the different parameters. Once we have the
posteriors for the various parameters we can characterize the
distributions in terms of the quantities of interest: expected
value, mode, standard uncertainty, probability intervals.

The above analysis has been performed for each cluster to derive
$T(z)$. We present the results in terms of $T(z)$ in Table
\ref{tab:tcmb}, where we give expected values and standard
deviations. Note that $T(z)$ at the
%redshift of cluster
cluster redshift
is
independent of the particular scaling assumed for the temperature
(i.e. the Lima model in this case); the only assumption made is
that the frequency obeys the standard scaling, $\nu(z) =
\nu_{0}(1+z)$ is valid. Rather, the $\alpha$ value is strictly
related to the specific model under consideration.

\begin{figure}
\centering
\includegraphics[width=8.5cm,keepaspectratio,angle=90]{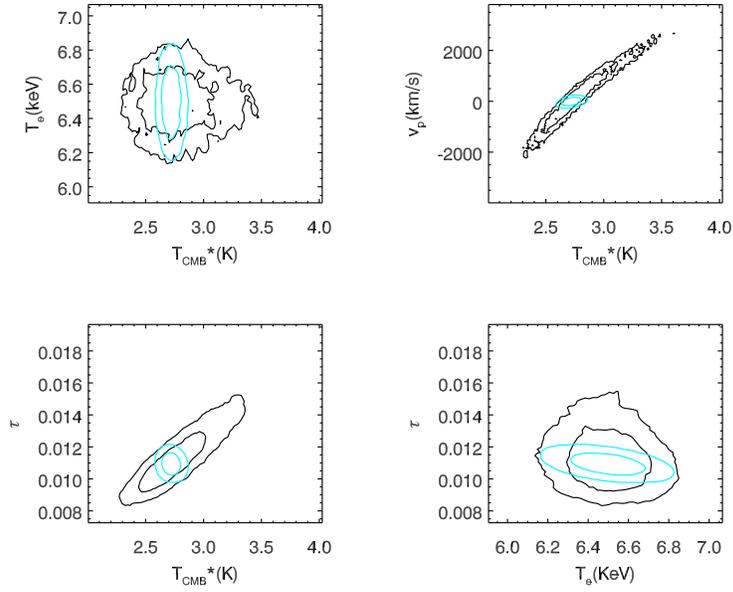}
\caption{Parameter correlations for a simulated cluster: the
contours show the $68\%$ and $95\%$ confidence limits from the
marginalized distributions. Black contours are obtained allowing
for a peculiar velocity prior with vanishing mean and standard
deviation 1000 km/s; cyan contours are obtained allowing for a
peculiar velocity prior with vanishing mean and standard deviation
100 km/s. $T^{\ast}_{\rm CMB}= T(z)/(1+z)$
}\label{Fig:contourplot}
\end{figure}

By studying the correlations between the variables, we checked
that the main degeneracies are between $T(z)$ and $\tau$, and
$T(z)$ and $\beta$. In particular, given the actual level of low
precision S-Z measurements, only the correlation between $T(z)$
and $\beta$ is always evident. We verified this behaviour by
simulating a dataset for a single cluster. We checked that, in
order to reduce the impact of the degeneracy between $T(z)$ and
$\beta$ and then to reduce the uncertainty in the determination of
$T(z)$, better knowledge of the peculiar velocity is required (see
Fig. (\ref{Fig:contourplot})). We used two different priors for
the peculiar velocity: one gaussian with vanishing mean and
standard deviation $1000$ km/s and the other gaussian with
vanishing mean and standard deviation $100$ km/s.

To obtain $\alpha$ we have performed a fit of the $T(z)$ data
points. Of course, since the distributions of $T(z)$ for
individual clusters are in general slightly non-gaussian and in
addition they are frequently skewed, performing a best fit as if
they were gaussian introduces a bias in the result. Nevertheless
we tested it is a good first approximation. The prior for $\alpha$
is flat in the range $\alpha\in[0,1]$, to account for the
theoretical constraints of the model \citep{lima2000}.

\subsection{Results}

To obtain the results we report below we assumed the flat prior
$\alpha\in[0,1]$. Employing the RI approach we deduce the most
probable value $\alpha=0.024^{+0.068}_{-0.024}$; all errors in
this section are at 68\% confidence. As discussed earlier, the use
of intensity ratios necessarily introduces bias in the inferred
parameter, although we attempted to minimize the bias by selecting
the measurement with the smallest fractional error to be in the
denominator of the ratios for each cluster. The more precise the
measurements, the weaker is this bias.
To estimate the bias in this result we repeated the calculation
with observational errors in values of $\Delta I$ (used in the
denominators) reduced to very small (relative) levels, thereby
isolating the impact of this systematic bias. Doing so yields
$\alpha=0.038^{+0.057}_{-0.038}$. Thus, when compared with the
value deduced from the actual data, $\alpha=0.024^{+0.068}_{-0.024}$,
a bias level of $\sim -0.014$ is indeed a relatively small fraction
of the statistical uncertainty, $\sim 0.057$.

\begin{figure}[h]
\centering
\includegraphics[width=8.5cm,keepaspectratio,angle=90]{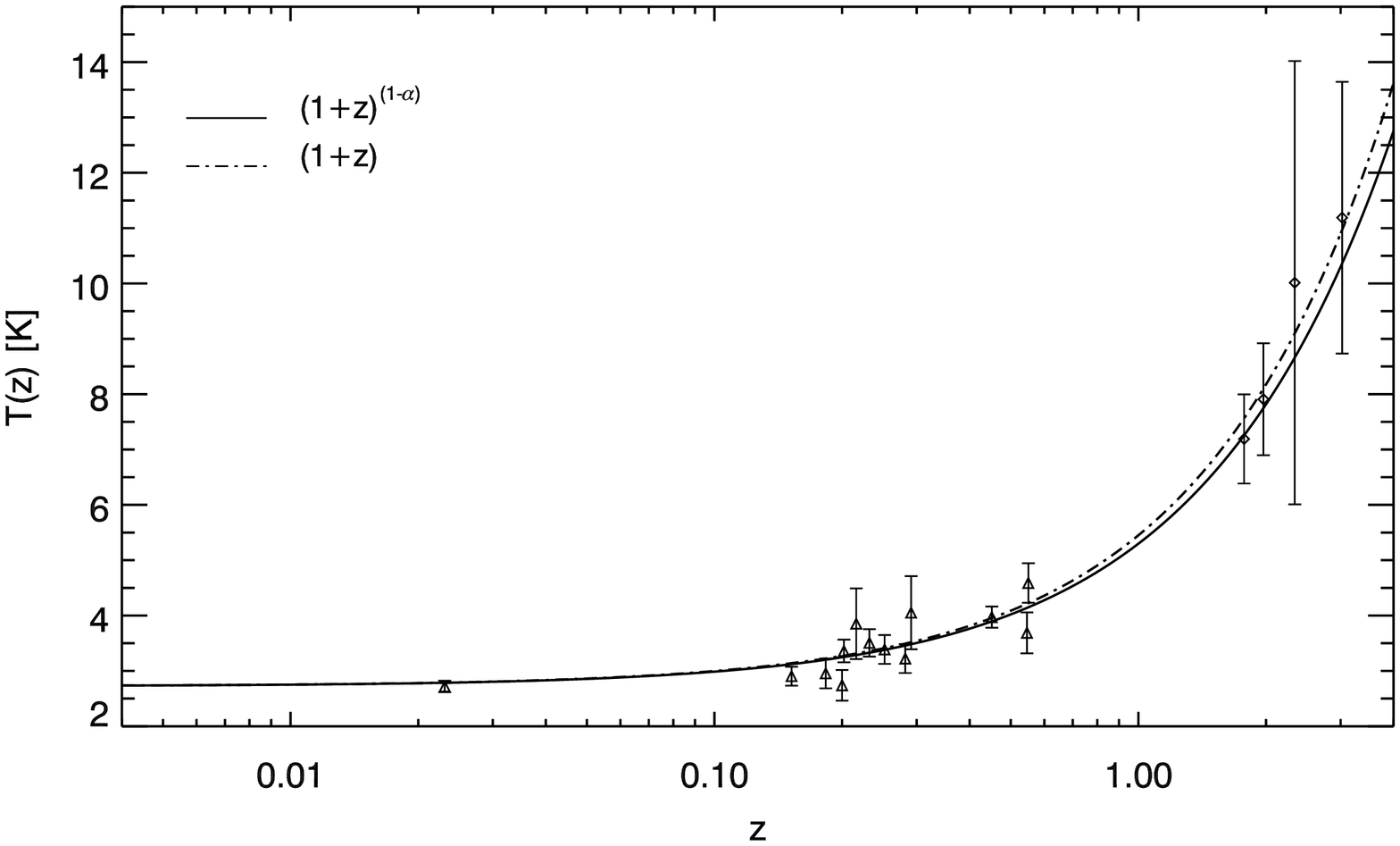}
\caption{$T(z)$ as deduced from S-Z spectra from complete data set
of clusters ($\vartriangle$) together with results deduced from
line transitions observations ($\lozenge$): \citep{cui2005,
ge1997,srianand2000,molaro2002}. The solid line is the best-fit to
the Lima scaling. The dot-dashed line is the standard scaling.}
\label{Fig:_righe}
\end{figure}

\begin{figure}[h!b]
\includegraphics[width=8.5cm,keepaspectratio]{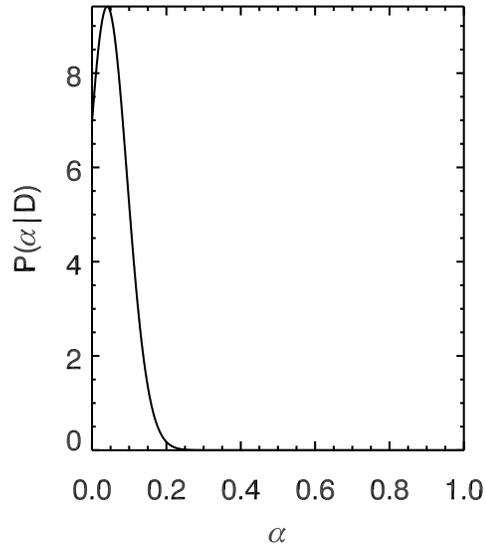}
\caption{Posterior of the $\alpha$ parameter, as obtained by
performing a fit of the $T(z)$ data points.} \label{Fig:alphap}
\end{figure}

The analysis was repeated employing the DI approach, adopting a
flat $\tau$ prior in the interval [0, 0.05]. In the first direct
3D integrations treatment we obtain $\alpha=0.026^{+0.033}_{-0.026}$.
The final alpha value we get by fitting the $T(z)$ data points obtained
with the MCMC treatment is $\alpha = 0.062 ^{+0.055}_{-0.062}$. In
order to check the efficiency of this method we have also considered
the $T(z)$ as deduced from line transitions observations (mainly
estimated by UV observations of interstellar clouds) (see Fig.
\ref{Fig:_righe}). The improvement on the constraints on $\alpha$
is not substantial in spite of the wider redshift range (as compared
to the S-Z data). The final alpha value we get by fitting the $T(z)$
data points is $\alpha = 0.041^{+0.038}_{-0.041}$. Due to the shape
of the posterior (see Fig. \ref{Fig:alphap}) as a consequence of the
strong prior on $\alpha$, it is more meaningful to present this
result as just an upper limit, i.e. $\alpha \leq 0.079$ at $68\%$
probability level.

Our three different treatments yield consistent results. More
%importantly,
important, no significant evidence is found for a deviation from
the redshift dependence of the CMB temperature predicted in the
standard model.

In principle, the intensity-ratio approach to determine $T(z)$
would seem preferable over the DI method since dependence on
$\tau$ is only second order, as compared with first order linear
dependence on this quantity, which therefore
%need
needs to be
marginalized over in the DI method. This is so even if the former
method is somewhat biased due to the arbitrariness in selecting
the intensity change used in the denominator of the intensity
ratios.
Our RI upper limit, $\alpha \leq 0.092$ at the 68\% confidence
level, is weaker than that obtained in the DI analysis, $0.059$;
this reflects the fact that the 68\% confidence region is wider
than the usual $1\sigma$ uncertainty region when the distribution
is non-gaussian, such as the one we get with the RI approach.
With more precise measurements, or - equivalently - a larger number of
clusters, the distribution of ratios would approach a gaussian and would
therefore be less affected by this bias. Ongoing and near future
surveys, including those planned with ACT, APEX-SZ, SPT, and Planck,
as well as detailed S-Z mapping of a sample of nearby clusters with
the MAD \cite{lamagna2002} and balloon-borne OLIMPO \cite{masi2003}
experiments, will provide much more precise and uniform datasets
that will largely remove the bias in the RI approach. While there
is no such bias in the DI approach, the marginalization over values
of $\tau$ lowers the usefulness of this approach. On the other hand,
spatially resolved spectroscopic observations of galaxy clusters (as
proposed with the SAGACE satellite, a small mission project approved
for Phase A by the Italian Space Agency) would allow for breaking the
degeneracy between $T(z)$ and cluster parameters.

\section{Summary}\label{sec:summary}

By adopting a more general redshift scaling law that includes the
prediction of the standard model when $\alpha=0$, we are able to
test both adiabatic and non-adiabatic models. While we obtain
somewhat different results for $\alpha$ in the above three
treatments, the degree of overlap of the respective uncertainty
intervals implies that these values are statistically consistent.
Already with the current sample of 13 clusters with medium-quality
S-Z data, we are able to verify the standard scaling law at a good
level of precision. More precise measurements of the $T(z)$
scaling law could possibly have interesting ramifications on
setting constraints on the variation of fundamental constants over
cosmological time.

We wish to thank G. D'Agostini for useful discussions and the
referee for helpful suggestions. MS gratefully acknowledges useful
discussions with Ran Shimon. This work has been supported in part
by MIUR/PRIN 2006 (prot.2006020237) and University of Rome, La
Sapienza (Ateneo prot.C26A0647AJ).

\appendix

\section*{Appendix A: Ratio Distribution}

We briefly summarize some of the basic elements of the theory of ratio
distributions employed in this work. If two variables $x$ and $y$ are
drawn from distribution functions $S(x)$ and $T(y)$, respectively, then
the probability that their ratio is in the interval $[r,r+dr]$ is
\begin{eqnarray}
\tilde{P}(r<x/y<r+dr)=\int_{x=-\infty}^{\infty}\int_{y=rx}^{(r+dr)x}S(x)T(y)dx\ dy,
\end{eqnarray}
which can be shown to imply that
\begin{eqnarray}
P(r)=\int_{-\infty}^{\infty}|x|S(x)T(rx)dx
\end{eqnarray}
where in the last step we took the absolute value of the Jacobian so that the
distribution function is non-negative. For gaussian distributions $S$ and
$T$ and finite $r$, this integral can be carried out analytically.

%In the case of three independent data points $x$, $y$, $z$,
%gaussianly distributed, so dealing with two ratios $u=y/x$, and
%$v=z/x$ , where $u= U(x,y)$ and $v=V(x,z)$, the density function
%$f(u,v)$ is:
%\begin{equation}
%f(u, v)=\int \delta[u-U(x,y)] \delta[v-V(x,z)]f(x,y,z)dx dy dz
%\end{equation}
%where $f(x,y,z)= f(x)f(y)f(z)$ because data are independent by
%definition, and which once integrated by using the substitution
%$y= ux$, $z= vx$ gives
%\begin{equation}
%f(u,v) \propto \int x^2 e^{-\frac{(x-\rho_1)^2}{2\sigma_{1}^{2}}}
%e^{-\frac{(ux-\rho_2)^2}{2\sigma_{2}^{2}}}
%e^{-\frac{(vx-\rho_3)^2}{2\sigma_{3}^{2}}} dx
%\end{equation}
%(see for example \citealt{dagos2003b}).

In the case of three independent, gaussianly distributed,
data points $x$, $y$, $z$, and considering the two ratios $r_{1}=y/x$,
and $r_{2}=z/x$ , where $r_{1}= R_{1}(x,y)$ and $r_{2}=R_{2}(x,z)$,
the joint density function $P(r_{1},r_{2})$ is
\begin{equation}
P(r_{1}, r_{2})=\int \delta[r_{1}-R_{1}(x,y)]
\delta[r_{2}-R_{2}(x,z)]f(x,y,z)dx dy dz
\end{equation}
where $f(x,y,z)= S(x)T(y)U(z)$ because data are independent by
definition, and which once integrated, after employing
$y= r_{1}x$ and $z= r_{2}x$, gives
\begin{equation}
P(r_{1},r_{2}) \propto \int x^2 e^{-\frac{(x-\rho_1)^2}{2\sigma_{1}^{2}}}
e^{-\frac{(r_{1}x-\rho_2)^2}{2\sigma_{2}^{2}}}
e^{-\frac{(r_{2}x-\rho_3)^2}{2\sigma_{3}^{2}}} dx
\end{equation}
(see for example \citealt{dagos2003b}).

In the case of $n$ ratios of gaussianly distributed $n+1$
data-points we similarly obtain the probability for the nontrivial
$n$ ratios (in addition to the ratio $r_0$ which equals unity) to
be {\it simultaneously} $r_{1}$, $r_{2}$,....,$r_{n}$
\begin{eqnarray}
P(r_{1},r_{2},...,r_{n};
\theta,\beta_{r};\alpha)\propto\int_{x=-\infty}^ {\infty}dx\cdot
|x|^{n}\exp\left[-\sum_{i=1}^{n+1}\frac{(xr_{i-1}-\rho_{i})^{2}}
{2\sigma_{i}^{2}}\right] ,
\end{eqnarray}
where $\rho_{i}$ and $\sigma_{i}$ are the expectation values and widths of
the individual gaussians. In our case (Table 2) $n$ can be either 1, 2, or
3. We carried out these 1D integrations numerically. In practice, the ratios
$r_{i}$ are the theoretical ratios of SZ observations at the various frequency
bands which are functions of the gas temperatures $\theta$ and cluster
peculiar velocities $\beta_{r}$, as well as the CMB temperature or $\alpha$.
The values $\rho_{i}$ are measured with errors $\sigma_{i}$ (Table 2).
The next step is to marginalize
%Eq. (A3)
Eq. (A5)
over the gas temperature $\theta$
and cluster velocity $\beta_{r}$.

A generic problem of ratio distributions is their inherent bias, which can
be illustrated as follows. Assume two measurements $\rho_{1}$ and $\rho_{2}$
with gaussian errors $\sigma_{1}$ and $\sigma_{2}$, respectively. The
probability function for the ratio $r$ is
\begin{eqnarray}
P(r)=\frac{1}{2\pi\sigma_{1}\sigma_{2}}
\int_{x=-\infty}^{\infty}xe^{-\frac{(x-\rho_{1})^{2}}{2\sigma_{1}^{2}}}
e^{-\frac{(xr-\rho_{2})^{2}}{2\sigma_{2}^{2}}}dx.
\end{eqnarray}
It can be easily verified that this probability function is indeed normalized,
$\int_{-\infty}^{\infty}P(r)dr=1$. Similarly, we can calculate the expectation
of the ratio $r$; $\langle r\rangle=\int_{-\infty}^{\infty}P(r)rdr$. A change
of variables leads to
\begin{eqnarray}
\langle r\rangle=\frac{1}{2\pi\sigma_{1}\sigma_{2}}
\int_{-\infty}^{\infty}\frac{dx}{x}e^{-\frac{(x-\rho_{1})^{2}}{2\sigma_{1}^{2}}}
\int_{-\infty}^{\infty}z dz e^{-\frac{(z-\rho_{2})^{2}}{2\sigma_{2}^{2}}}.
\end{eqnarray}
The second integral can be readily calculated and yields
$\sigma_{2}\rho_{2}\sqrt{2\pi}$, and the first integral is
\begin{eqnarray}
\int_{-\infty}^{\infty}\frac{dx}{x}e^{-\frac{(x-\rho_{1})^{2}}{2\sigma_{1}^{2}}}
=\frac{1}{\rho_{1}}\int_{-\infty}^{\infty}\frac{dx}{1+\frac{x}{\rho_{1}}}
e^{-\frac{x^{2}}{2\sigma_{1}^{2}}}.
\end{eqnarray}
Assuming that the measurement at the denominator of the ratio is very
accurate, i.e. that $\sigma_{1}\ll\rho_{1}$, the denominator in the
integral can be expanded in powers of $\frac{x}{\rho_{1}}$, and the
integration is then carried out term by
term. The leading order expectation value for the ratio is then
\begin{eqnarray}
\langle r\rangle=\frac{\rho_{2}}{\rho_{1}}\left[1+\frac{1}{2}(\frac{\sigma_{1}}{\rho_{1}})
^{2}+O\left((\frac{\sigma_{1}}{\rho_{1}})^{4}\right)\right].
\end{eqnarray}
This demonstrates the bias in ratio distributions; it is dominated
by the error of the denominator and therefore we choose in all our
calculations the denominator to be the measurement with the
smallest fractional error as to minimize the bias. In higher
dimensional distribution functions the definition of the bias is
not unique
%, so it
and therefore
cannot be unambiguously calculated and corrected
for. Note, also, that the bias is in the expectation value of the
ratio of SZ temperatures, and not directly in $\alpha$. The exact
change in $\alpha$ as a result of this bias is difficult to
predict a priori, but for high sensitivity measurements with large
samples of clusters the effective fractional error drops
significantly, and the bias in $\alpha$ is much weaker. As in the
DI method our likelihood function is obviously non-gaussian in
$\alpha$ and we define the average $\alpha$ as the median value of
that distribution.

\np

\begin{table}[h]
\caption{Cluster redshift and gas temperatures} \label{tab:tableX}
\renewcommand{\footnoterule}{}  % to avoid a line before footnotes
\begin{minipage}[htbp]{\columnwidth}
\begin{tabular}{lcc}
\hline \hline Cluster & z  & $kT_e^{a}$ (keV)\\ \hline
  A1656 & 0.023&$8.25\pm0.10^{b}$\\
  A2204 & 0.152& $11.53\pm2.80$ \\
  A1689 & 0.183& $9.59\pm2.80$  \\
  A520 & 0.200& $8.33\pm0.76$ \\
  A2163 & 0.202& $16.18\pm3.86$ \\
  A773 & 0.216& $6.62\pm1.30$ \\
  A2390 & 0.232& $10.13\pm1.22$ \\
  A1835 & 0.252& $13.45\pm4.02$ \\
  A697 & 0.282& $10.60\pm1.13$\\
  ZW3146 & 0.291& $8.12\pm1.00$ \\
  RXJ1347 & 0.451& $13.69\pm2.64$ \\
  CL0016+16 & 0.546& $10.10\pm2.57$ \\
  MS0451-0305 & 0.550& $10.68\pm2.93$ \\
\hline
\multicolumn{3}{l}{\footnotesize{$^a$\cite{bonamente2006}}}\\
\multicolumn{3}{l}{\footnotesize{$^b$\cite{arnaud2001}}}\\
\end{tabular}
\end{minipage}
\end{table}

\begin{table}[h]
\caption{SZE measurements of 13 clusters by different experiments
expressed in central thermodynamic temperature.}
\label{tab:tableSZ} \footnotesize \centering
\renewcommand{\footnoterule}{}  % to avoid a line before footnotes
\begin{tabular}{lcccccc}
  \hline \hline
 & OVRO+BIMA$^a$ &  \multicolumn{4}{c}{SuZIE II$^b$} & SCUBA$^c$  \\
\cline{3-6}
  Cluster & $\Delta T_{30\rm GHz}$ &  $\Delta T_{145\rm GHz}$ & $\Delta T_{221\rm GHz}$ & $\Delta T_{273\rm GHz}$ & $\Delta T_{355\rm GHz}$ & $\Delta T_{353\rm GHz}$ \\
  & (mK) & (mK) & (mK) & (mK) & (mK) & (mK) \\
\hline
  A520  & $-0.66\pm0.09^d $ &  $-0.44\pm0.13$ & $0.14\pm0.14$ & -& $1.78\pm0.41$ & $2.60\pm0.57$  \\
  A697  & $-1.22\pm0.12$ &  $-0.93\pm0.13$ & $0.41\pm0.16$ & - & $2.80\pm0.62$ & - \\
  A773  & $-1.08\pm0.11$ &  $-0.91\pm0.16$ & $0.04\pm0.25$ &-& $2.40\pm0.89$ & $2.7\pm2.0$ \\
  A1689  & $-2.06\pm0.17$ &  - & - & - &- & $2.93\pm0.40$ \\
  A1835  & $-2.90\pm0.21$ &  $-1.74\pm0.26$ & $0.14\pm0.41$ & $1.73\pm0.59$ &-& -  \\
  A2204  & $-3.22\pm0.32$ &  $-0.65\pm0.10$ & $0.21\pm0.10$ & &$1.85\pm0.44$& - \\
  A2390  & -&  $-0.91\pm0.10$ & $-0.10\pm0.17$ &- & $1.23\pm0.34$& $2.40\pm0.44$ \\
  CL0016+16  & $-1.44\pm0.09$&  $-0.57\pm0.23$ & $0.66\pm0.46$ & $1.82\pm0.68$ & -& $1.96\pm0.64$  \\
  MS0451-03& $-1.48\pm0.09$ & $-0.779\pm0.065$ & $-0.21\pm0.10$ & $0.91\pm0.66$ & $1.16\pm0.34$ & - \\
  RXJ1347  & $-5.15\pm0.60$ &  $-3.22\pm0.39$ & $-0.10\pm0.39$ & - & $6.2\pm1.5$ & $5.36\pm0.54$ \\
  ZW3146 & $-2.02\pm0.25$ & $-1.56\pm0.39$ & $-0.25\pm0.48$ & - & $3.1\pm1.3$ & -  \\
  \hline
  \hline
  & OVRO$^d$ &  & MITO$^e$ & & & SCUBA$^c$  \\
\cline{3-5}
  Cluster  & $\Delta T_{32\rm GHz}$ &  $\Delta T_{143\rm GHz}$ & $\Delta T_{214\rm GHz}$ & $\Delta T_{272\rm GHz}$ &  & $\Delta T_{353\rm GHz}$ \\
  &(mK) &  (mK) & (mK) & (mK) &  & (mK) \\
  \hline
  A1656  & $-0.520\pm0.093$ & $-0.179\pm0.037$ & $0.033\pm0.080$ & $0.170\pm0.034$ &  & -\\
  \hline
  \hline
  & OVRO+BIMA$^a$ &  & SuZIE I$^f$ & & & SCUBA$^c$  \\
\cline{3-5}
  Cluster & $\Delta T_{30\rm GHz}$ &  $\Delta T_{142\rm GHz}$ & $\Delta T_{217\rm GHz}$ & $\Delta T_{268\rm GHz}$ &  & $\Delta T_{353\rm GHz}$ \\
  & (mK) &  (mK) & (mK) & (mK) & & (mK) \\
  \hline
  A2163 & $-1.89\pm0.17$ &  $-1.011\pm0.098$& $-0.21\pm0.16$ & $0.66\pm0.24$ &  & -\\
  \hline
  &&&&&&\\
\multicolumn{7}{l}{\footnotesize{$^a$\cite{bonamente2006};
 $^b$\cite{benson2003,benson2004};}}\\
 \multicolumn{7}{l}{\footnotesize{$^c$\cite{zemcov2007};
 $^d$\cite{her95,mason01};}}\\
  \multicolumn{7}{l}{\footnotesize{$^e$\cite{depetris2002,savini2003};$^f$\cite{holza97a}.}}\\
\end{tabular}
\end{table}

\begin{table}
\caption{$T(z)$ values as estimated for each cluster. We report
expected values and standard deviations.} \label{tab:tcmb}
\begin{minipage}[htbp]{\columnwidth}
\begin{tabular}{lcc}
\hline \hline Cluster & $T(z)$ \\
 &  (K)\\ \hline
     A1656 & 2.72 $\pm$  0.10  \\
     A2204 & 2.90 $\pm$  0.17  \\
     A1689 & 2.95 $\pm$  0.27  \\
      A520 & 2.74 $\pm$  0.28  \\
     A2163 & 3.36 $\pm$  0.20  \\
      A773 & 3.85 $\pm$  0.64  \\
     A2390 & 3.51 $\pm$  0.25  \\
     A1835 & 3.39 $\pm$  0.26  \\
      A697 & 3.22 $\pm$  0.26  \\
    ZW3146 & 4.05 $\pm$  0.66  \\
   RXJ1347 & 3.97 $\pm$  0.19  \\
 CL0016+16 & 3.69 $\pm$  0.37  \\
    MS0451 & 4.59 $\pm$  0.36  \\
\hline
\end{tabular}
\end{minipage}
\end{table}


\begin{thebibliography}{}

\bibitem[Arnaud {\it et al.} 2001]{arnaud2001} {Arnaud} M.,
 {Aghanim} N., {Gastaud} R., {Neumann} D. M., {Lumb} D., {Briel}
 U., {Altieri} B., {Ghizzardi} S., {Mittaz} J., {Sasseen} T. P.
 and {Vestrand} W. T., 2001, {A\&A} 365, L67-L73


\bibitem[Battistelli \ea 2002]{battiste2002}
{Battistelli} E.~S., {De Petris} M., {Lamagna} L., {Melchiorri}
F., {Palladino} E., {Savini} G., {Cooray} A., {Melchiorri} A.,
{Rephaeli} Y., {Shimon} M., 2002, {ApJL} 580, L101

\bibitem[Benson \ea 2003]{benson2003}
{Benson} B.~A., {Church} S.~E., {Ade} P.~A.~R., {Bock} J.~J.,
{Ganga} K.~M., {Hinderks} J.~R., {Mauskopf} P.~D., {Philhour} B.,
{Runyan} M.~C., {Thompson} K.~L., 2003, {\apj} 592, 674-691


\bibitem[Benson \ea 2004]{benson2004}
{Benson} B.~A., {Church} S.~E., {Ade} P.~A.~R., {Bock} J.~J.,
{Ganga} K.~M., {Henson} C.~N., {Thompson} K.~L., 2004, {\apj} 617,
829-846


\bibitem[Bonamente \ea 2006]{bonamente2006}
{Bonamente} M., {Joy} M.~K., {LaRoque} S.~J., {Carlstrom} J.~E.,
{Reese} E.~D., {Dawson} K.~S., 2006, {\apj} 647, 25-54


\bibitem[Combes and Wiklind 1999]{combes1999}
{Combes} F. and {Wiklind} T., 1999, ASP Conf. Ser. 156: Highly
Redshifted Radio Lines, 210, ed. by Carilli C. L., Radford S. J.
E., Meten  K. M. \& Langston G. I., San Francisco: ASP (1999),
Proc. Conf. Green Bank, WV, USA, 9-11 October


\bibitem[Cui \ea 2005]{cui2005}
{Cui} J., {Bechtold} J., {Ge} J., {Meyer} D.~M., 2005, {\apj} 633,
649-663

\bibitem[D'Agostini 2003]{dagos2003}
{D'Agostini} G., 2003, {Rep. Prog. Phys.} 66, 1383-1419

\bibitem[D'Agostini 2003b]{dagos2003b}
{D'Agostini} G., 2003b, {Bayesian reasoning in data analysis - A
critical introduction}, {World Scientific Publishing}

\bibitem[De Gregori \ea 2008]{degregori2008}
{De Gregori} S., {Conte} A., {De Petris} M.,{Lamagna} L., {Luzzi}
G., {Battistelli} E.S., {Savini} G., 2008, {Il Nuovo Cimento della
Societ�Italiana di Fisica}, ISSN: 1826-9877.

\bibitem[De Petris {\it et al.} 2002]{depetris2002}
{De Petris} M., {D'Alba} L., {Lamagna} L., {Melchiorri} F.,
{Orlando} A., {Palladino} E., {Rephaeli} Y., {Colafrancesco} S.,
{Kreysa} E., and {Signore} M., 2002, {ApJ}, 574, L119-L122

\bibitem[Fabbri$,$ Melchiorri \& Natale 1978] {fabbri1978}
{Fabbri} R. and {Melchiorri}. F. , \& {Natale} V. 1978, Astrophysics
and Space Science, 59, 223

\bibitem[Ge \ea 1997]{ge1997}
{Ge} J., {Bechtold} J., {Black} J.~H., 1997, {\apj} 474, 67-+

\bibitem[Gelman \ea 1992]{gelman1992}
{Gelman} A., {Rubin} D., 1992, {Statistical Science} 7, 457

\bibitem[Hastings 1970]{hastings1970}
{Hastings} W.K., 1970, {Biometrika} 57, 97-109

\bibitem[Herbig {\it et al.} 1995]{her95}
{Herbig} T., {Lawrence} C.R. and {Readhead} A.C.S., 1995, {ApJ}
449, 5

\bibitem[Holzapfel {\it et al.} 1997a]{holza97a}
{Holzapfel} W. L., {Wilbanks} T. M., {Ade} P. A. R., {Church} S.
E., {Fischer} M. L., {Mauskopf} P. D., {Osgood} D. E. and {Lange}
A. E., 1997a, {ApJ} 479, 17


\bibitem[Itoh$,$ Kohyama \& Nozawa 1998]{itoh1998}
{Itoh} N., {Kohyama} Y. and {Nozawa} S., 1998, {ApJ} 502, 7

\bibitem[Itoh {\it et al.} 2002]{itoh2002}
{Itoh} N., {Sakamoto} T., {Kusano} S., {Kawana} Y. and {Nozawa}
S., 2002, {A\&A} 382, 722-729


\bibitem[Lamagna \ea 2002]{lamagna2002}{Lamagna} L., {De Petris} M.,
{Melchiorri} F., {Battistelli} E. S., {De Grazia} M., {Luzzi} G.,
{Orlando} A., {Savini} G., 2002, {AIP} Conference Proceedings,
616, 92, editors: M. De Petris and M. Gervasi


\bibitem[Lewis and Bridle 2002]{lewis2002}
{Lewis} A., {Bridle} S., 2002, {Phys. Rev. D} 66, 103511

\bibitem[Lima {\it et al.} 2000]{lima2000}
{Lima} J. A. S., {Silva} A. I. and {Viegas} S.M., 2000, {MNRAS}
312, 747-752

\bibitem[LoSecco {\it et al.} 2001]{losecco2001}
{LoSecco} J. M., {Mathews} G. J. and {Wang} Y., 2001, {Phys. Rev.
D} 64, 123002

\bibitem[Masi {\it et al.} 2003]{masi2003}
{Masi}, S., {Ade}, P., {de Bernardis}, P., {Boscaleri}, A., {De
Petris}, M., {De Troia}, G., {Fabrini}, M., {Giacometti}, M.,
{Iacoangeli}, A., {Lamagna}, L., {Lange}, A., {Lubin}, P.,
{Mauskopf}, P., {Melchiorri}, A., {Melchiorri}, F., {Nati}, F.,
{Nati}, L., {Orlando}, A., {Pascale}, E., {Piacentini}, F.,
{Pierre}, M., {Polenta}, G., {Rephaeli}, Y., {Romeo}, G., {Yvon},
D. 2003, {Memorie della Societa Astronomica Italiana}, 74, 96+-

\bibitem[Mason {\it et al.} 2001]{mason01}
{Mason} B.S., {Myers} S.T. and {Readhead} A.C.S., 2001, {ApJL}
555, L11


\bibitem[Mather {\it et al.} 1999]{mather1999}
Mather, J. C., Fixsen, D. J., Shafer, R. A., Mosier, C. and
Wilkinson, D. T., 1999, {ApJ} 512, 511-520

\bibitem[Matyjasek 1995]{matyjasek1995}
{Matyjasek}, 1995, {Phys. Rev. D}, 51, 4154-4159

\bibitem[Metropolis {\it et al.} 1953]{metropolis1953}
{Metropolis} N., {Rosenbluth} A. W., {Rosenbluth} M.N., {Teller}
A.H., {Teller} E., 1953, {Journal of Chemical Physics} 21,
1087-1092

\bibitem[Molaro {\it et al.} 2002] {molaro2002}
{Molaro} P., {Levshakov} S. A., {Dessauges-Zavadsky} M. and
{D'Odorico} S., 2002, {A\&A} 381, L64-L67


\bibitem[Opher \& Pelison 2005]{opher2005}
{Opher} R., {Pelison} A., 2005, {Mon. Not. R. Astron. Soc.} 362,
167-170

\bibitem[Overduin \& Cooperstock 1998]{overduin1998}
{Overduin} J.M., {Cooperstock} F.I., 1998, {Phys. Rev. D} 58,
043506

\bibitem[Puy 2004]{puy2004}
Puy D., 2004, {A\& A} 422, 1-9

\bibitem[Reese {\it et al.} 2002]{reese2002}
{Reese} E.~D., {Carlstrom} J.~E., {Joy} M., {Mohr} J.~J., {Grego}
L., {Holzapfel} W.~L., 2002, {\apj} 581, 53-85

\bibitem[Rephaeli 1980]{reph1980}{Rephaeli} Y., 1980, {ApJ}
241, 858


\bibitem[Rephaeli 1995]{reph95}{Rephaeli} Y. 1995, {ApJ}
445, 33


\bibitem[Savini {\it et al.} 2003]{savini2003}
{Savini}, G., {Orlando}, A., {Battistelli}, E.~S., {de Petris},
M., {Lamagna}, L., {Luzzi}, G., {Palladino}, E. 2003, {New
Astronomy} 8, 727


\bibitem[Shimon and Rephaeli 2004]{shimon2004}
Shimon, M., \& Rephaeli, Y. 2004, New Astronomy, 9, 69

\bibitem[Srianand {\it et al.} 2000]{srianand2000}
{Srianand} R., {Petitjean} P., {Ledoux} C., 2000, {\nat} 408,
931-935


\bibitem[Sunyaev \& Zel'dovich 1972]{SunZel72}
Sunyaev R.A. \& Zel'dovich Ya. B., 1972,  Comm. Astrophys. Space
Phys., 4, 173

\bibitem[Sunyaev \& Zel'dovich 1980]{SunZel80}
Sunyaev R.A. \& Zel'dovich Ya. B., 1980, MNRAS {\bf 190}, 413


\bibitem[Verde 2007]{verde2007}
{Verde} L., 2007, {ArXiv e-prints 0712.3028}, 712

\bibitem[Zemcov \ea 2007]{zemcov2007}
{Zemcov} M., {Borys} C., {Halpern} M., {Mauskopf} P., {Scott} D.,
2007, {\mnras} 376, 1073-1098

\end{thebibliography}
\end{document}